# Direct Observation of Vortices and Antivortices Generation in Phase-Separated Superconductor Sn-Pb Solder

H. Arima[1]*, T. Murakami[2], Y. Kinoshita[3], H. Sepehri-Amin[4], M. Tokunaga[3], T. Nojima[5]*, Y. Mizuguchi[2]*

[1]*National Metrology Institute of Japan, National Institute of Advanced Industrial Science and Technology, Tsukuba, 305-8565, Japan.*
[2]*Department of Physics, Tokyo Metropolitan University; Hachioji, 192 - 0397, Japan.*
[3]*Institute for Solid State Physics, University of Tokyo, Kashiwa, 277-8581, Japan.*
[4]*National Institute for Materials Science; Tsukuba, 305-0047, Japan.*
[5]*Institute for Materials Research, Tohoku University, Sendai 980-8577, Japan.*

*Corresponding author: h-arima@aist.go.jp, t.nojima@tohoku.ac.jp, mizugu@tmu.ac.jp

**Abstract**

**Quantised vortices in type-II superconductors provide insights into the mechanisms of superconductivity; however, the generation of antivortices, which is characterised by magnetisation antiparallel to the external magnetic field, remains less understood. In this study, we investigate a Sn-Pb solder, which is a superconductor with phase-separated Sn and Pb phases, and we report the observation of both vortices and antivortices. Scanning superconducting quantum interference device microscopy revealed the presence of both vortices and antivortices, while magneto-optical imaging demonstrated flux avalanches. Our results indicate that Sn in the Sn-Pb solder behaves as a type-II superconductor when magnetic fluxes are trapped despite bulk Sn being a type-I superconductor with a transition temperature ($T_c^{Sn}$) of 3.7 K. Our findings suggest that the size and proximity effects with Pb synergistically contribute to inducing type-II superconductivity in Sn. Vortices were observed at temperatures as high as 5 K, exceeding the bulk $T_c^{Sn}$. Moreover, the interplay between the type-I superconducting Pb phase and type-II superconducting Sn phase generates antivortices, thereby providing a mechanism to accommodate excess magnetic flux. This study highlights new research on composites combining type-I and type-II superconductors.**

Quantised vortices with a magnetic flux $\Phi_0$ of $2.07 \times 10^{-15}$ Wb, which can be generated by the penetration of magnetic flux in a superconductor, are a well-established feature of type-II superconductors. The theoretical basis for magnetic field behaviour in type-II superconductors was established by Abrikosov, who predicted that vortices would form lattices, now known as Abrikosov lattices [1]. Subsequently, this prediction was experimentally validated using techniques such as neutron diffraction [2], the bitter method [3], and scanning tunnelling microscopy (STM) [4]. These vortices provide valuable insight into the mechanism of superconductivity. Although ordinary vortices have been well studied, research on antivortices characterised by magnetisation antiparallel to the external field remains limited.

Various approaches to form antivortices in superconductors have been investigated. One such approach involves using small superconductors with specific geometric structures. It has been theoretically predicted that vortex–antivortex pairs can form in small superconductors because of confinement effects [5,6]. Experiments have demonstrated the formation of antivortices in superconductors where antidots are introduced into square thin-film samples [7]. Another approach for antivortex generation involves strong electron correlations. For example, in the iron chalcogenide thin-film superconductor Fe(Se,Te), antivortices have been observed without an external magnetic field [8], induced by spin-orbit coupling between the magnetic moments of impurity Fe atoms and superconducting current. Similarly, in an iron-based thin-film superconductor $EuFe_2(As_{0.79}P_{0.21})_2$, in which superconductivity coexists with weak ferromagnetism, antivortices are formed through interactions between the magnetic domains and superconducting state [9]. Most studies on antivortices focus on small superconductors with dimensions that can be compared to the coherence length ($\xi$) or magnetic penetration depth ($\lambda$). Recently, antivortices have been reported in bulk superconductors, such as $UTe_2$, which is a chiral superconductor that breaks time-reversal symmetry [10].

In this paper, we report on Sn-Pb solder as a new material in which antivortices are generated. Unlike previously studied systems, the bulk Sn-Pb solder is a conventional material without magnetic interactions, which is typical of strongly correlated electron systems.

Recently, nonvolatile magneto-thermal switching was observed in superconducting Sn-Pb solders below the superconducting transition temperature of Pb ($T_c^{Pb}$) [11]. A Sn-Pb solder is an alloy with phase separation between the Sn and Pb regions. The solder exhibited low thermal conductivity immediately after zero-field cooling (ZFC) because of the formation of Cooper pairs. However, the thermal conductivity increases when a magnetic field is applied because of the transition to the normal state, and it remains high even after the magnetic field is removed. This nonvolatile magneto-thermal switching is driven by the magnetic flux trapping within the superconducting state after the application of a magnetic field. Specific heat measurements revealed that $T_c^{Pb}$ and $T_c^{Sn}$ are consistent with their bulk values immediately after ZFC, where $T_c^{Sn}$ represents a superconducting transition temperature of Sn. However, when the magnetic field is reduced to zero after field cooling (FC), the specific heat results show that $T_c^{Pb}$ remains unchanged, while $T_c^{Sn}$ is suppressed. This implies that the nonvolatile magneto-thermal switching originates from the coexistence of the bulk superconducting Pb phases and the weakened superconducting (or normally conducting) Sn phases because of magnetic flux trapping. Given that both Sn and Pb are type-I superconductors that expel magnetic flux below their critical fields, this assumption remains hypothetical. Flux trapping in Sn-Pb solders was first reported in the 1960s [12-16]; however, the origin of this flux trapping still remains unclear because both Sn and Pb are type-I superconductors. We conducted surface observations of the Sn-Pb solder in its superconducting state using scanning superconducting quantum interference device (SQUID) microscopy to further investigate flux trapping in this solder. Contrary to earlier assumptions, our results demonstrate that Sn in the Sn-Pb solder behaves as a type-II superconductor, enabling the trapping of magnetic flux as vortices. Remarkably, antivortices were generated simultaneously.

Figure 1 illustrates the selected scanning SQUID microscopy (SSM) images in the superconducting state of the Sn50-Pb50 solder with a Sn-to-Pb mass ratio of 50:50. All images are shown in Figures S1 and S2 in the Supplementary Information. The SSM images presented in Figure 1 depict the trapped magnetic flux at a zero magnetic field after FC from temperatures ($T$) above $T_c^{Pb}$ to 5.0 K in various FC fields. Figure 1(a) shows the SSM image obtained immediately after ZFC with a scanning range of 300 μm × 300 μm. Despite the ZFC condition, slight positive and negative magnetic fluxes are observed, which can be attributed to the residual geomagnetic field. The contrast between the regions of positive and negative magnetic fluxes becomes more pronounced with an increase in the FC field (Figures 1(b–e)). Localised magnetic flux areas emerge, indicating concentrated magnetic flux trapping. High-resolution SSM images with a scanning range of 200 μm ×200 μm are shown in Figures 1(f–j) to further investigate these regions. The distinction between the positive and negative magnetic flux regions becomes clearer with an increase in the FC field and a more concentrated flux is observed in certain localised areas. Zoomed-in views of one such region, marked by black squares in Figure 1(g), are shown in Figures 1(k) and 1(l). These images reveal isolated flux quanta, with area integrations yielding values of $0.98\Phi_0$ in Figure 1(k) and $-1.02\Phi_0$ in Figure 1(l), which is consistent with quantised vortices. This implies that the presence of a type-II superconducting region in the solder, where both vortices and antivortices are generated. A three-dimensional (3D) mapping of the magnetic flux is shown in Figures 1(m-1), 1(m-2), 1(n-1), and 1(n-2) to further characterise the magnetic flux distribution. Figure 1(m-1), which corresponds to Figure 1(g), reveals multiple spike-like regions in the positive flux areas, signifying the presence of multiple vortices. In contrast, as shown in Figure 1(m-2), the area of negative magnetic flux contains a few spike-like shapes, which are indicative of antivortices. Figure 1(n-1), corresponding to Figure 1(j), shows multiple vortices in the positive flux areas. Figure 1(n-2) indicates that the number of spike-like shapes increases with increasing FC fields, suggesting that higher fields promote the generation of antivortices.

The SSM images of the solder in the superconducting state were obtained for the Sn90-Pb10 solder with a Sn-to-Pb mass ratio of 90:10 and the Sn10-Pb90 solder with a Sn-to-Pb mass ratio of 10:90 to further investigate the conditions under which vortices and antivortices were generated in the solder (Figure 2). All images are shown in Figures S3 and S4 in the Supplemental Information. The SSM images of Sn90-Pb10 are displayed in Figures 2(a–e), measured at a zero magnetic field after FC from $T > T_c^{Pb}$ to 4.5 K. Figure 2(a) shows that almost no magnetic flux was trapped immediately after ZFC. However, distinct regions of positive and negative trapped magnetic fluxes became evident with an increase in the FC field. No localised magnetic flux regions were observed for Sn90-Pb10, which is in contrast to the behaviour observed for Sn50-Pb50. Both the positive and negative magnetic flux regions in Sn90-Pb10 exhibit uniform distributions. However, the SSM images of Sn10-Pb90 exhibited different trends from those of the Sn90-Pb10 and Sn50-Pb50 solders. Figure 2(f) shows that, immediately after the ZFC, a small amount of magnetic flux was trapped owing to the residual geomagnetic field. The amount of trapped magnetic flux increased with an increase in the FC field, and the contrast between the regions of positive and negative magnetic fluxes became more pronounced. However, the behaviour of the trapped magnetic flux varies across different regions. Figures 2(g–j) show that a localised magnetic flux appeared in the lower-right region, whereas no such localisation was observed in the upper-left region. To investigate this in greater detail, we focused on the upper left and lower right regions of the 3 μT FC image in Figure 2(h). Vortices are present in the region shown in Figure 2(k), which is similar to the behaviour observed for the Sn50-Pb50 solder. In contrast, the flux was uniformly distributed in the region shown in Figure 2(l), which resembled the behaviour observed for Sn90-Pb10. These observations suggested that the Sn-to-Pb mixing ratio plays a crucial role in determining the presence of type-II superconducting regions within the Sn-Pb solder.

Possible candidates for the type-II superconducting regions in Sn-Pb solder include (1) the alloy phase of Sn and Pb, (2) Pb phase, and (3) Sn phase. The elemental composition of the solder was analysed using scanning electron microscopy with energy-dispersive X-ray spectroscopy (SEM-EDS) and scanning transmission electron microscopy (STEM) with energy-dispersive X-ray spectroscopy (STEM-EDS) to investigate these possibilities. The results are presented in Extended Data Figure 1. The solder samples were wire-shaped, and therefore, elemental mapping was performed on both circular and rectangular cross sections for all compositions. No directional dependence was observed in the distribution of the Sn and Pb phases (Extended Data Figures 1(a), 1(b), 1(d), 1(e), 1(h), and 1(i)). In the Sn50-Pb50 solder, Sn and Pb were homogeneously mixed ((Extended Data Figures 1(a) and 1(b)). The SEM-EDS line profile further confirmed that Sn and Pb were completely phase-separated (Extended Data Figure 1(c)). For Sn10-Pb90, Sn was dispersed within a Pb-rich matrix (Extended Data Figures 1(d) and 1(e)). The STEM-EDS line profile confirmed complete phase separation between Sn and Pb (Extended Data Figures 1(f) and 1(g)). Conversely, in Sn90-Pb10, the Pb phase is dispersed within the Sn-rich matrix (Extended Data Figures 1(h) and 1(i)). Although the Sn-rich phase is completely phase separated, the Pb phase contains ~3 at. % Sn in an alloyed state, as indicated by the STEM-EDS mapping and line profile (Extended Data Figures 1(j) and 1(k)). Alloys are typically type-II superconductors, and therefore, vortex formation is expected in the Sn90-Pb10 solder. However, as shown in Figure 2, vortices were not observed in the Sn90-Pb10 solder. This absence of vortices in the Pb phase of Sn90-Pb10 can be attributed to the formation of a superconducting Sn-Pb alloy phase as filaments, which prevented vortex formation. In contrast, vortices were observed in Sn50-Pb50 and Sn10-Pb90, where Sn and Pb were completely phase separated. This indicates that either the Sn or Pb phase behaved as a type-II superconductor in these compositions.

The type of superconductivity is determined by the relationship between $\lambda$ and $\xi$. A superconductor is classified as type-I if $\sqrt{2}\lambda<\xi$ and as type-II if $\sqrt{2}\lambda>\xi$. In thin film samples, where the sample size is comparable to $\lambda$ and/or $\xi$, samples that are bulk type-I superconductors exhibit type-II superconductivity. The critical thickness separating type-I and type-II superconductivity is ~2.5 μm for Pb and ~1.5 μm for Sn [17]. As shown in Extended Data Figure 1, the sizes of the Sn and Pb phases in the Sn50-Pb50 solder are on the order of a few micrometres. This implies that both the Pb and Sn phases in the solder have the potential to transition into type-II superconductors.

To this end, we examined the superconducting state of Pb. Extended Data Figure 2 shows the temperature dependence of the specific heat ($C$) for Sn90-Pb10, Sn50-Pb50, and Sn10-Pb90, as well as the critical magnetic field of Pb ($H_c^{Pb}$) estimated from specific heat measurements. Further, to investigate the specific heat in the flux-trapped state, $C(H)$ was measured at various magnetic fields after 150 mT FC from $T > T_c^{Pb}$ to 1.8 K. The specific heat shown in Extended Data Figure 2 was calculated as $(C(H)-C(150\ \text{mT}))/T$, where $C(150\ \text{mT})$ corresponds to the specific heat of Sn and Pb in the normal state. This calculation emphasised the contribution of the electronic specific heat of Pb. For Sn90-Pb10, as shown Extended Data Figure 2(a), specific heat jumps corresponding to the superconducting transitions of Pb at 7.2 K and Sn at 3.7 K were observed at 0 mT. For Sn50-Pb50 and Sn10-Pb90a, as shown in Extended Data Figures 2(b) and 2(c), a specific heat jump caused by the superconducting transition of Pb at 7.2 K was observed at 0 mT; however, the superconducting transition of Sn was not detected. In all solders, $T_c^{Pb}$ decreased with an increasing applied magnetic field. Extended Data Figures 2(d–f) present $H_c^{Pb}$ values for each solder determined from the specific heat measurements. The $H-T$ phase diagrams were fitted using the equation $H_c(T) = H_c(0)(1-(T/T_c)^2)$, which yield $\mu_0 H_c^{Pb}(0)$ values of 78 mT for Sn90-Pb10, 77.5 mT for Sn50-Pb50, and 85 mT for Sn10-Pb90, where $\mu_0$ represents a vacuum permeability. For type II superconductors, the upper critical magnetic field ($H_{c2}$) typically exceeds the critical field ($H_c$) of type I superconductors. In Pb thin films, $H_{c2}$ for type-II superconducting Pb exceeds 100 mT for films thinner than 0.5 μm [18]. However, the $H_c$ values estimated from this experiment were consistent with those of bulk Pb, supporting its classification as a type-I superconductor. This assumption was corroborated by magnetisation measurements.

Extended Data Figure 3 presents the $M-T$ and $M-H$ curves for the Sn10-Pb90, Sn50-Pb50, and Sn90-Pb10 solders, where $M$ represents the magnetisation. The $M-T$ curve for Sn10-Pb90, shown in Extended Data Figure 3(a), indicates that the ZFC and FC curves coincide from $T_c^{Pb}$ to the irreversible temperature ($T_{irr}$) of 5.5 K. Below $T_{irr}$, the magnetisation becomes irreversible because of differences between the ZFC and FC conditions. This suggests the coexistence of type-I and type-II superconducting regions in the Sn10-Pb90 solder. In Extended Data Figure 3(d), the $M-H$ curve immediately after ZFC to 2.0 K shows that $M$ behaves reversibly below $H_c^{Pb}$; however, it becomes irreversible below 20 mT ($H_{irr}$) owing to flux trapping. At 5.0 K, as shown in Extended Data Figure 3(g), the magnetisation is predominantly influenced by Pb, and the $M-H$ curve shows a reversible behaviour below $H_c^{Pb}$. For Sn50-Pb50, ZFC and FC exhibited irreversible behaviours below $T_c^{Pb}$, as shown in Extended Data Figure 3(b). Extended Data Figures 3(e) and 3(h) show the magnetic hysteresis characteristics of type-II superconductors. At 5.0 K, as shown in Extended Data Figures 3(h), a reversible magnetic field range exists below $H_c^{Pb}$; however, this range is narrower than that of Sn90-Pb10. The magnetic properties of Sn10-Pb90 were similar to those of

Sn50-Pb50. Extended Data Figure 3(c) shows that ZFC and FC exhibited irreversible behaviour below $T_c^{Pb}$. The magnetisation curves in Extended Data Figures 3(f) and 3(i) exhibited magnetic hysteresis. At 5.0 K, Extended Data Figure S11(i) reveals a magnetic field range in which the magnetisation exhibited slightly reversible behaviour below $H_c^{Pb}$. From the above observations, we concluded that Pb remains a type-I superconductor in the Sn-Pb solder. Therefore, the type-II superconducting region in the solder was most likely associated with the Sn phase. In solder studies, examining the electromagnetic properties of Sn as a superconductor is challenging because the superconducting current of Pb shields the solder below $T_c^{Pb}$. Consequently, direct evidence supporting the classification of Sn as a type-II superconductor remains limited. An indirect piece of evidence suggesting that Sn is a type-II superconductor is self-heating in the solder [19]. The solder exhibited anomalous heating when the specific heat was measured after FC. Based on the observation of the vortices in the solder, the self-heating phenomenon was likely caused by the motion of the vortices. This heating occurred below 4 K–$T_c^{Sn}$, thereby indicating that the phase involved in the vortex formation was most likely Sn. It is important to emphasise that magnetic-flux trapping can play a crucial role in the emergence of type-II superconductivity in Sn. The M–H curve of the Sn50-Pb50 solder at 2 K immediately after ZFC shown in Extended Data Figure 3(e) indicates that the magnetisation initially decreased linearly with an increasing magnetic field, which is consistent with perfect diamagnetism. At ~30 mT, the magnetic flux began to penetrate the solder near the critical field of Sn, thereby deviating from the linear decrease in magnetisation. If Sn is a type-II superconductor, flux penetration occurs at a lower critical field, below 30 mT. Thus, we propose that magnetic flux penetration is a key factor that contributes to the type-II superconducting behaviour of Sn.

Figure 3 illustrates a schematic of a magnetic flux line looping through the solder. As shown in Figure 3(a), when $T>T_c^{Pb}$ or $H>H_c^{Pb}$, both Sn and Pb are in the normal state, which enables the external magnetic field to penetrate the entire solder uniformly. However, vortices are trapped when FC is performed from $T>T_c^{Pb}$ to $T<T_c^{Pb}$ at $H>H_c^{Pb}$ and the magnetic field is reduced to zero ($H<H_{c2}^{Sn}$), where $H_{c2}^{Sn}$ represents the upper critical field of Sn. According to the basic principles of electromagnetism, magnetic flux lines originating from these vortices must form closed loops. In conventional superconductors, these flux lines loop around the outside of the sample, with the magnetic flux entering from one side of the sample and exiting from the other, thereby creating a closed flux loop. However, the situation is different for the solders. The magnetic flux originating from the vortices in the Sn region is surrounded by Pb in a type-I superconducting state, which prevents the penetration of the magnetic flux. Consequently, the flux lines cannot loop externally around the solder. To resolve this situation, the magnetic flux lines are confined within the solder, looping through regions near the magnetic penetration depths of Sn and Pb $\lambda_{Sn}+\lambda_{Pb}$, where $\lambda_{Sn}$ and $\lambda_{Pb}$ represent the magnetic penetration depths of Sn and Pb, respectively. Re-considering the SSM images of Sn50-Pb50 3 μT FC, as shown in Figures 1(b) and 1(g), reveals not only positive vortices but also a uniform region of negative magnetic flux extending across the entire image. This uniform region corresponds to magnetic flux penetrating within $\lambda_{Sn}+\lambda_{Pb}$. However, if the total trapped flux exceeds the capacity of $\lambda_{Sn}+\lambda_{Pb}$, it can no longer be fully confined within this region. The solder generates antivortices as an additional strategy to accommodate the excess flux and maintain closed flux loops. Extended Data Figure 4 illustrates the FC field dependence of the total trapped magnetic flux in the SSM images of the Sn50-Pb50, Sn90-Pb10, and Sn10-Pb90 solders. For the Sn90-Pb10 and Sn10-Pb90 solders, the trapped magnetic flux increased proportionally with an increasing FC field. In contrast, for the Sn50-Pb50 solder, a magnetic flux was induced in the direction opposite to that of the FC field. This behaviour can be attributed to Sn50-Pb50, which contained a higher proportion of type-II superconducting Sn than that of the other solders, thereby allowing it to accommodate the excess magnetic flux in the form of antivortices.

The magnetic flux trapping of the Sn50-Pb50 solder exhibited distinct characteristics compared to those of the other solders. However, in the SSM experiments for Sn50-Pb50, the lowest temperature was limited to 5.0 K, and the highest applied magnetic field was 20 μT. Magneto-optical microscopy experiments were conducted for investigating magnetic flux trapping at lower temperatures and higher magnetic fields. Figure 4(a) shows the $B$–$H$ curve ($B = 4\pi M + H$) obtained immediately after ZFC to 2.0 K. In the sample used for MO imaging, flux penetration began at ~10 mT. Around $H = H_c^{Pb}$, $B$ was equal to $H$, and as $H$ returned to zero, $B$ remained finite because of flux trapping. Figures 4(b–q) show selected MO images that correspond to the magnetic field values shown in Figure 4(a). A complete set of MO images is shown in Figure S5 in the Supplementary Information. The MO image obtained immediately after ZFC at 2 K (Figure 4(b)) shows a uniform magnetic field distribution. However, negative magnetisation caused by perfect diamagnetism was observed when a 10 mT field was applied (Figure 4(c)). With an increase in the external magnetic field, the magnetic flux begins to penetrate the sample in a branched pattern near the edges. The entire solder transitioned to the normal state, thereby resulting in a uniform colour across the image. When the magnetic field was reduced from $H>H_c^{Pb}$, a trapped magnetic flux became apparent in the entire solder at 20 mT (Figure 4(i)). At zero magnetic field (Figure 4(k)), a further reduction in the applied field revealed that the magnetic flux was still trapped within the solder. The magnetic flux exits the solder in a branched pattern. As the field continued to decrease, additional flux exited the solder, and eventually, the entire solder transitioned to the normal state, thereby yielding a uniform image. Figure S5 in the Supplemental Information shows the results for a magnetic field swept from negative to positive values. These results mirror the aforementioned observations, with the magnetic flux trapped in the negative direction exiting the solder in a similar branched pattern. The branched patterns observed in the MO images resemble a flux avalanche [20,21], which occurs when an external magnetic field is locally enhanced at the edges of a sample, causing the magnetic flux to move abruptly. Such flux avalanches are also evident in the temperature-dependent ZFC measurements shown in Figure S6, as well as in the temperature dependence after FC, as shown in Figure S7. Thus, vortices are present even at high magnetic fields and temperatures lower than those investigated in the SSM experiment. A magnetic flux opposite the external magnetic field is not induced. In the MO image of the solder at 0 mT after 100 mT, as shown in Fig. 4 (l), the region where the magnetic flux was trapped was a uniformly positive magnetic flux in the same direction as the external magnetic field. In other words, the antivortex was not induced by an excessively high external magnetic field. Therefore, it is possible that the magnetic flux line that loops the solder differs between cases with low and high magnetic fields. In this study, there is no experimental result that analogises the loop of the magnetic flux line in the high magnetic field; however, it is considered that it passes through the vortex and $\lambda_{Pb}+\lambda_{Sn}$.

      We found that Sn in the solder behaves as a type-II superconductor, with vortices observed directly using SSM and indirectly through flux avalanches using MO microscopy. However, two questions remain unresolved: (1) mechanism by which Sn transitions to a type-II superconducting state and (2) observation of vortices at 5.0 K in SSM, despite the bulk $T_c^{Sn}$ being 3.7 K. One possible explanation is the size effect. The SSM results reveal almost no vortices in the Sn90-Pb10 solder with a large Sn phase, whereas vortices are observed in the Sn10-Pb90 solder with a small Sn phase. In addition, Sn nanowires have been reported to exhibit an enhanced $T_c^{Sn}$ of 5.5 K, although their dimensions differ from those of Sn in Sn-Pb solders [22]. Although the size effect appears to be essential in the Sn-Pb solder, it alone cannot fully explain the superconducting behaviour of Sn. For example, in the Sn10-Pb90 solder, in which the Sn size is smaller than that in the Sn50-Pb50 solder, the SSM images reveal a bias in regions where vortices are present, thereby suggesting that another mechanism is involved. Another possible mechanism involves the proximity effect of Pb. This effect allows the superconducting order parameters to leak from Pb into Sn, thereby enabling Sn to acquire superconductivity. In the Sn50-Pb50 solder, the Sn

phase is relatively large; however, the extensive contact area between Sn and Pb may enable a significant leakage of the superconducting order parameter. Conversely, in the Sn10-Pb90 solder, the Sn-Pb interface may be too small to limit the effect of the proximity effect. However, the range of the proximity effect in a superconductor-metal junction is limited to 1 µm [23,24], whereas the Sn phase in the solder exceed this scale. This indicates that the proximity effect alone is insufficient for explaining the superconductivity of Sn. Instead, it is plausible that the size and proximity effects act synergistically, which enable Sn to transition to a type-II superconducting state with an enhanced $T_c^{Sn}$. Furthermore, the trapped magnetic flux plays a crucial role in this transition. The superconducting properties of the Sn-Pb solder after ZFC resemble those of type-I superconductors, which implies that flux trapping is essential for the emergence of type-II behaviour.

In this study, we investigated phase-separated superconductors in Sn-Pb solders by focusing on the magnetic flux trapping mechanism. Sn was identified as a type-II superconductor within the solder using SSM and MO microscopy. Vortices were directly observed in the SSM images, whereas flux avalanches indicative of quantised vortices were detected using MO microscopy. Despite the intrinsic type-I superconductors Sn and Pb, our findings provide clear evidence that the Sn phase exhibits type-II superconductivity. The mechanism by which Sn transitions to type-II superconductivity remains partially understood; however, it is likely a combined effect of the size and proximity effects with Pb when the magnetic flux is trapped. In addition, antivortices were observed in the Sn-Pb solder. This highlights the unique interplay between the type-I superconducting Pb phase and type-II superconducting Sn phase within the solder, which facilitates the formation of closed flux loops and antivortices. These findings not only advance our understanding of the superconducting properties of Sn-Pb solder but also provide new insights into the coexistence of type-I and type-II superconductivity in heterogeneous systems. The generation of vortices and antivortices in such materials has potential implications for applications requiring controlled flux trapping and magnetic flux manipulation.

## Methods

### Samples

The solders used in this study, namely, Sn90-Pb10, Sn50-Pb50, and Sn10-Pb90, were manufactured by Sasaki Solder Industry CO. Ltd., with Sn: Pb mass ratios of 90:10, 50:50, and 10:90, respectively. All solders were flux-core-free with a diameter of 1.5 mm.

### Characterisation

The composition of the solder surface was analysed using SEM-EDS (Carl ZEISS CrossBeam 1540EsB) and STEM-EDS (FEI Titan G2 80-200).

### Scanning SQUID microscopy

Magnetic images of the solder were obtained using a commercially available scanning SQUID microscope (SQM-2000, SII NanoTechnology). The sensor chip used in this experiment consisted of a dc-SQUID element made of Nb thin film and a pickup coil with a diameter of 10 μm at its tip. The measurement temperature was 4.5 K for Sn90-Pb10 and 5.0 K for Sn50-Pb50 and Sn10-Pb90. FC was performed at various magnetic fields above the superconducting transition temperature of Pb, with the magnetic field reduced to zero at the measurement temperature. The geomagnetic field during the experiment was −3 μT, and external magnetic field adjustments were made accordingly. For example, in measurements at 10 μT, FC was performed at 7 μT, and the external magnetic field was adjusted to −3 μT. The scanning range was 300 μm × 300 μm with a 2.5 μm step size, and 200 μm × 200 μm with a 2 μm step size. For Sn10-Pb90, additional measurements were conducted over scanning ranges of 130 μm × 130 μm and 140 μm × 140 μm with a 2 μm step.

### Magneto-optical imaging

In the magneto-optical imaging experiment, a polariser film was attached to the sample, and the magnetic field response was observed using magneto-optical microscopy [25]. The temperature and magnetic field were controlled using a physical property measurement system (PPMS, Quantum Design), into which an infinity-corrected objective lens was inserted. The results presented in this paper were normalised at 2.0 K image $I(H, 2.0\text{ K})$ using the temperature difference image between the superconducting solder at 2.0 K and the normal solder at 8.0 K, i.e. $(I(H, 2.0\text{ K}) - I(H, 8.0\text{ K}))/I(H, 2.0\text{ K})$.

**Physical property measurements**

Specific heat measurements were conducted using the relaxation method with the PPMS. The samples were mounted on the sample stage using Apiezon N grease. All specific heat measurements were conducted after FC. Previous studies reported that the Sn-Pb solder shows anomalous self-heating when the sample is slightly warmed during specific heat measurements [19]. Self-heating occurred only during the first measurement at each temperature and diminished in subsequent measurements. In this study, a third measurement was conducted to eliminate the effect of self-heating. Magnetisation measurements were conducted in the vibrating sample magnetometer (VSM) mode of a magnetic property measurement system (MPMS3, Quantum Design) equipped with a SQUID.

**Acknowledgements**

We thank T. Yagi, O. Miura, A. Yamashita, and N. Kurata for their support with the experiments and fruitful discussion on the results. This work was partly supported by JST-ERATO (JPMJER2201) and JSPS KAKENHI (Grant number 24K23040).

**Author contributions**

H.A., T.N., and Y.M. planned and supervised the study. H.A., Y.K., H.S.A., M.T., T.N., and Y.M. designed the experiments. H.A., T.M., Y.K., H.S.A., M.T., T.N., and Y.M. collected and analysed the data. H.A., T.N., and Y.M. prepared the manuscript. All authors discussed the results, developed the explanation of the experiments, and commented on the manuscript.

**Competing interest declaration**

The authors declare no competing financial interests.

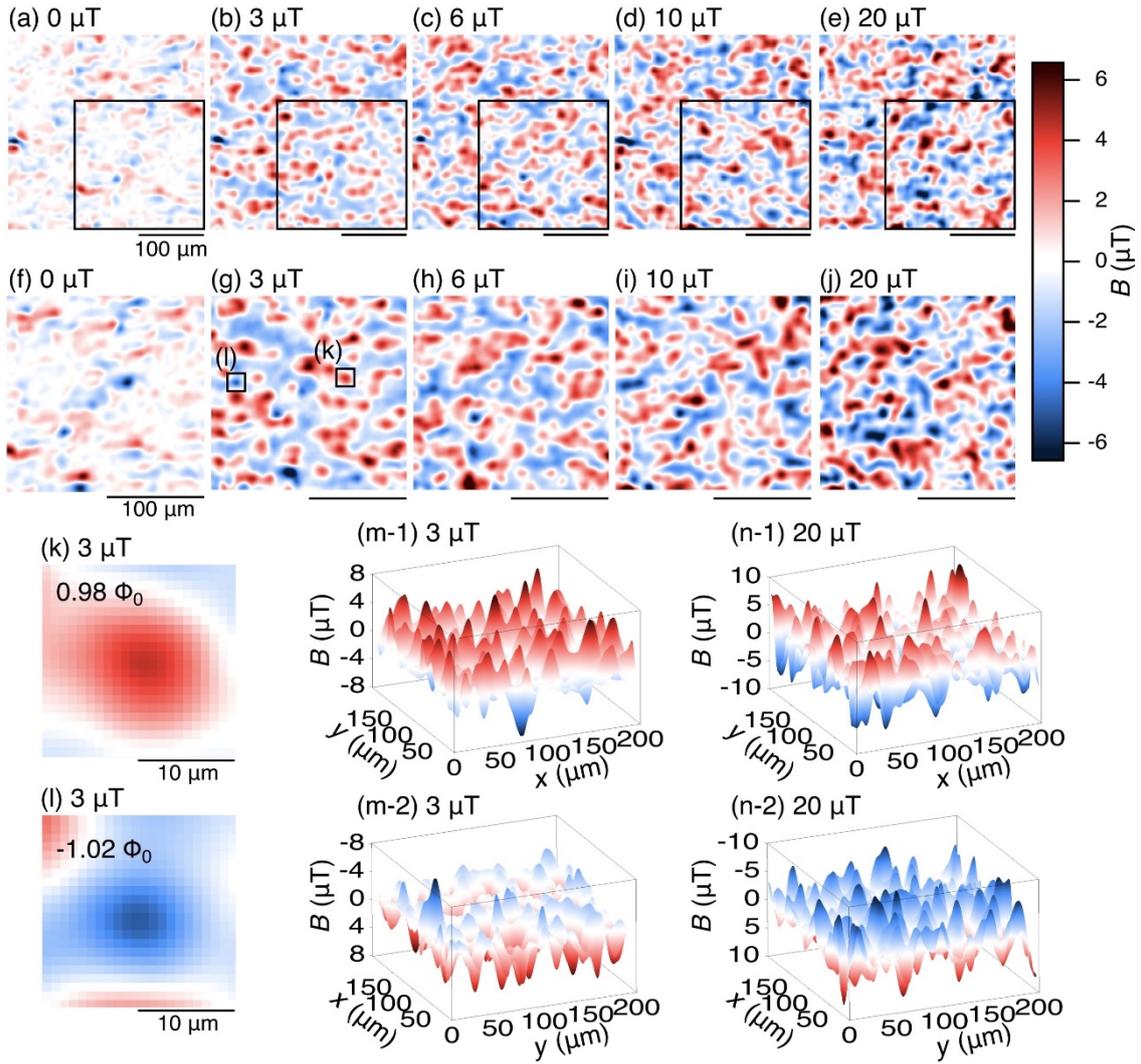

Fig. 1 SSM images of Sn50-Pb50 solders after FC. (a–e) SSM images at zero magnetic field after FC from $T > T_c^{Pb}$ to 5.0 K, with applied FC fields indicated above each image. The scanning range is 300 μm × 300 μm. (f–j) High-resolution SSM images corresponding to the black squares in (a)–(e), with a scanning range of 200 μm × 200 μm. (k), (l) Enlarged SSM images of isolated vortices obtained at zero magnetic field after 3 μT FC. Positive (k) and negative (l) flux images correspond to vortices and antivortices, respectively, with flux values close to $\pm\Phi_0$. (m-1), (m-2) 3D magnetic field mappings for image (g), showing (m-1) $B>0$ and (m-2) $B<0$ flux regions. (n-1), (n-2) 3D magnetic field mappings for image (j), obtained at zero magnetic field after 20 μT FC, displaying (n-1) $B>0$ and (n-2) $B<0$ flux regions. Scale bars apply to all images.

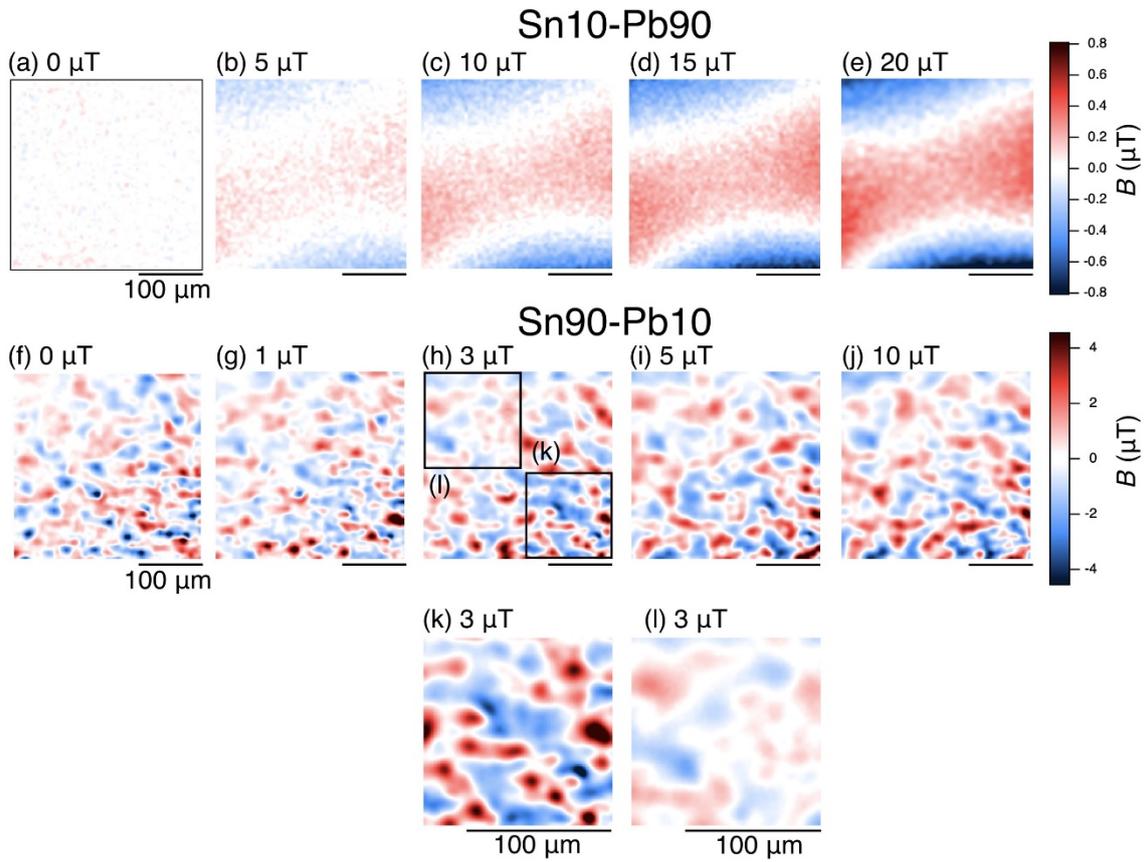

Fig. 2 SSM images of Sn90-Pb10 and Sn10-Pb90 solders after FC. (a–e) SSM images of Sn90-Pb10 obtained at zero magnetic field after FC from temperatures above $T_c^{Pb}$ to 4.5 K, with FC fields indicated above each image. (f–j) SSM images of Sn10-Pb90 obtained under the same conditions, but with a final temperature of 5.0 K. (k), (l) High-resolution images of the regions marked by black squares in (h), with scanning ranges of 140 μm × 140 μm for (k) and 130 μm × 130 μm for (l). Scale bars apply to all images.

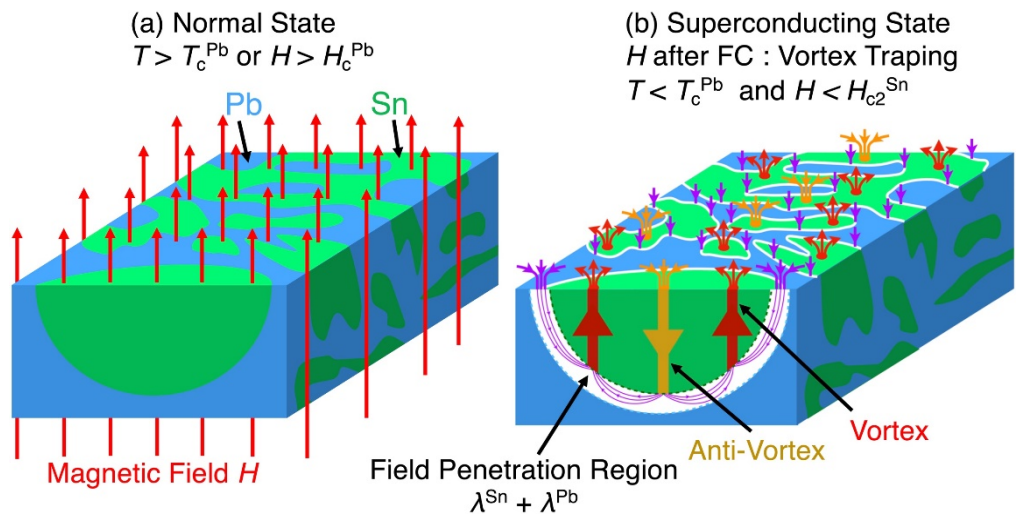

Fig. 3 Schematic of flux trapping and vortex formation in the Sn-Pb solder before and after the superconducting transition. (a) Magnetic flux distribution in the normal state, where both Sn and Pb are non-superconducting ($T > T_c^{Pb}$ or $H > H_c^{Pb}$), thereby enabling external magnetic fields to penetrate the whole sample. (b) Magnetic flux distribution when the solder transitions to the superconducting state ($T < T_c^{Pb}$ and $H < H_{c2}^{Sn}$) following FC. Vortices (red arrows) and antivortices (yellow arrows) are trapped in the superconducting regions, and flux lines (purple arrows) penetrate the solder up to the combined penetration depth ($\lambda_{Sn}+\lambda_{Pb}$).

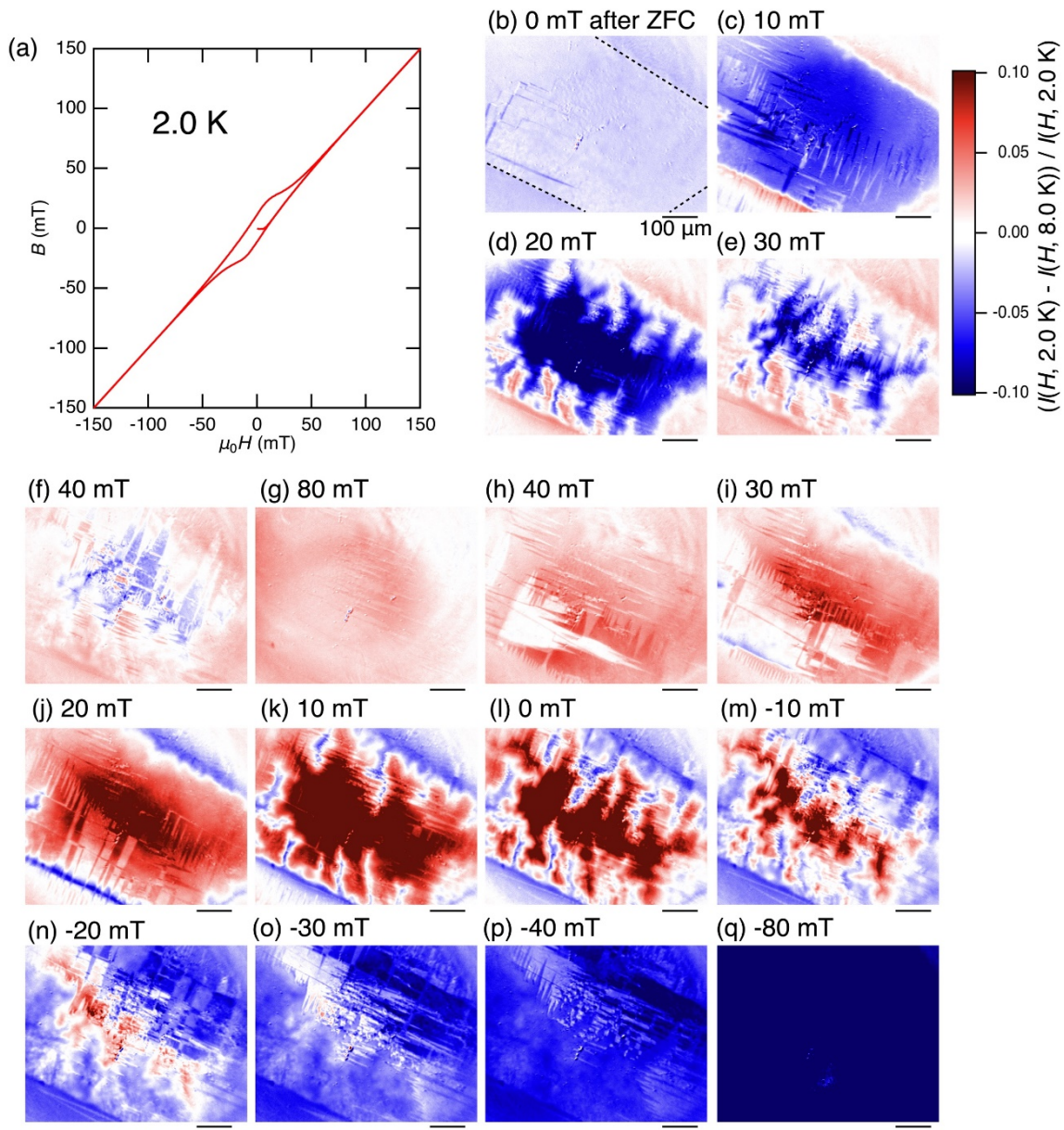

Fig. 4 Magnetic field dependence of flux behaviour in the Sn50-Pb50 solder observed via MO imaging. (a) $B$–$H$ curve measured immediately after ZFC to 2.0 K. (b–q) MO images corresponding to $B$ at different magnetic fields as indicated in (a). (b–g) illustrate flux penetration with an increase in the external magnetic field from 0 mT to 80 mT. (h–q) show the flux evolution with a decrease in the magnetic field from 80 mT to −80 mT. The colour scale represents the relative change in magnetic flux distribution normalised to the field at 2.0 K.

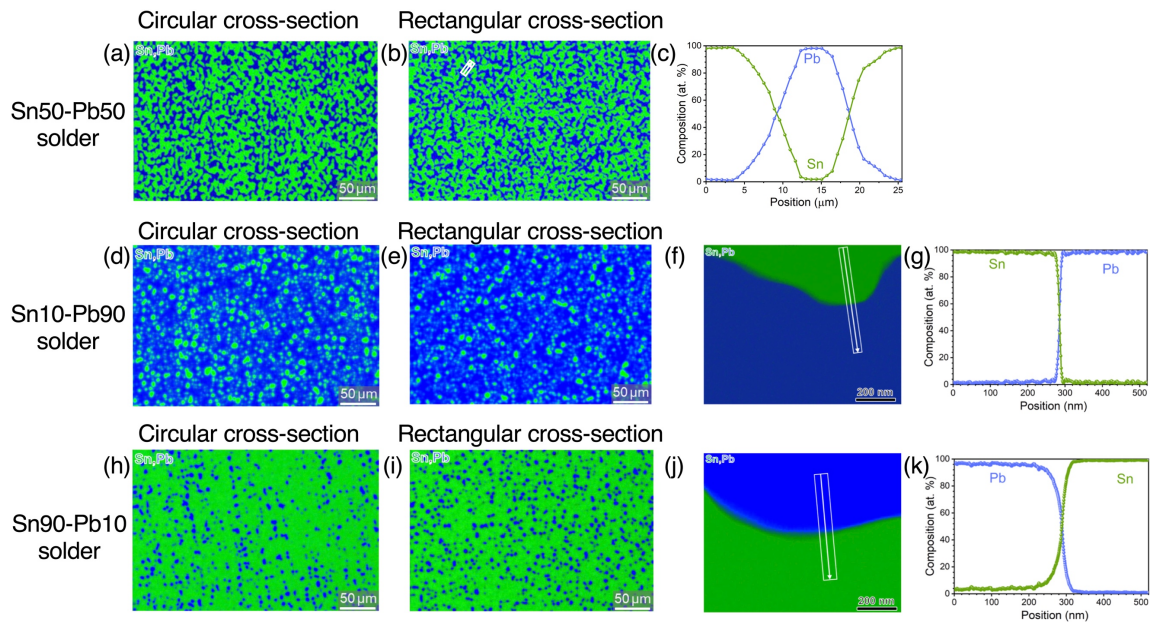

Extended Data Fig. 1 SEM-EDS and STEM-EDS analyses of Sn-Pb solder wires that show the elemental distribution of Sn and Pb in different cross-sectional views. (a), (b) SEM-EDS mapping images of (a) circular and (b) rectangular cross sections of the Sn50-Pb50 solder. (c) Line profile of the elemental composition along the arrow in (b). (d), (e) SEM-EDS mapping images of (d) circular and (c) rectangular cross-sections of the Sn10-Pb90 solder. (f) STEM-EDS mapping image of the Sn10-Pb90 solder. (g) Line profile of the elemental composition along the arrow in (f). (h), (i) SEM-EDS mapping images of circular (h) and rectangular (i) cross sections of the Sn90-Pb10 solder. (j) STEM-EDS mapping of the Sn90-Pb10 solder. (k) Line profile of the elemental composition along the arrow in (j).

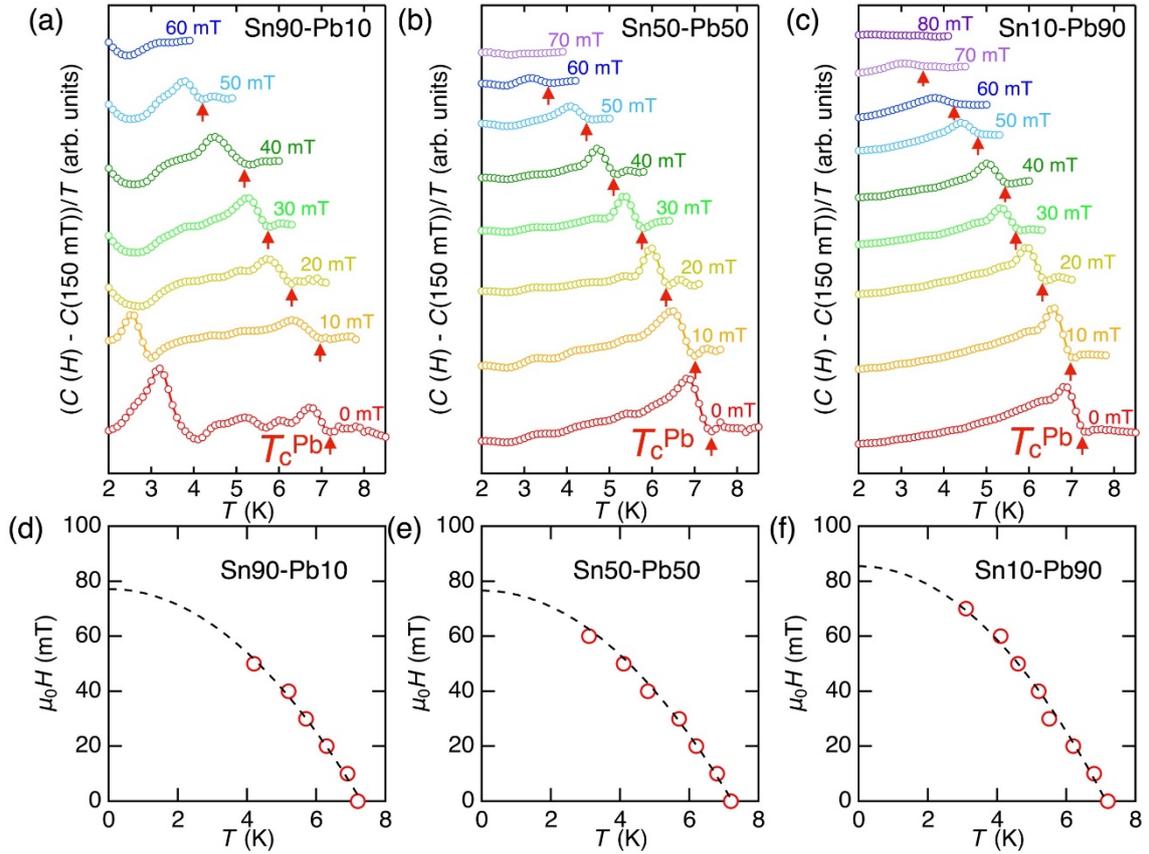

Extended Data Fig. 2 Temperature dependence of the specific heat and critical field in the Sn-Pb solder. (a)–(c) Temperature dependence of the specific heat for (a) Sn90-Pb10, (b) Sn50-Pb50, and (c) Sn10-Pb90 solders. The specific heat values on the vertical axis were calculated as $(C(H)-C(150\,\text{mT}))/T$, where $C(H)$ represents the specific heat at a magnetic field $H$, and $C(150\,\text{mT})$ represents the specific heat at 150 mT. The red arrows indicate $T_c^{Pb}$ for each sample. (d–f) Temperature dependence of the critical field estimated from the specific heat data for (d) Sn90-Pb10, (e) Sn50-Pb50, and (f) Sn10-Pb90. Dashed lines represent the theoretical fit.

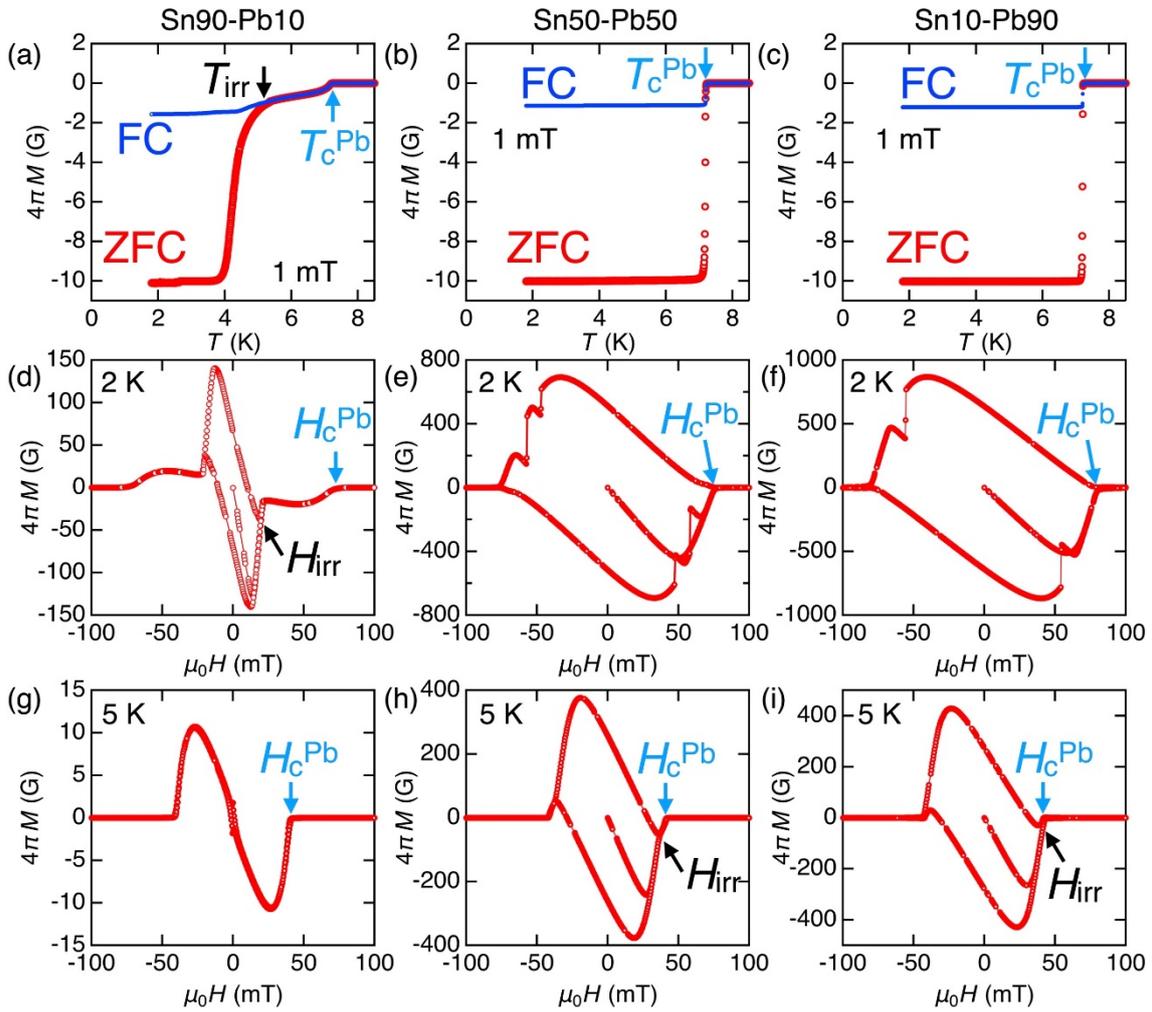

Extended Data Fig. 3 Temperature and field dependence of magnetisation in the Sn-Pb solder. (a–c) Temperature dependence of magnetisation for the (a) Sn90-Pb10, (b) Sn50-Pb50, and (c) Sn10-Pb90 solders measured under an applied field of 1 mT. The red curves correspond to ZFC measurements, whereas the blue curves represent FC measurements. $T_c^{Pb}$ is indicated by blue arrows and $T_{irr}$ is marked in (a). (d–f) Magnetic hysteresis curves measured at 2 K for the (d) Sn90-Pb10, (e) Sn50-Pb50, and (f) Sn10-Pb90 solders. $H_c^{Pb}$ is indicated by the blue arrows, whereas $H_{irr}$ is indicated in (d). (g–i) Magnetic hysteresis curves of the (g) Sn90-Pb10, (h) Sn50-Pb50, and (i) Sn10-Pb90 solders at 5 K. $H_c^{Pb}$ and $H_{irr}$ are indicated by the blue and black arrows, respectively.

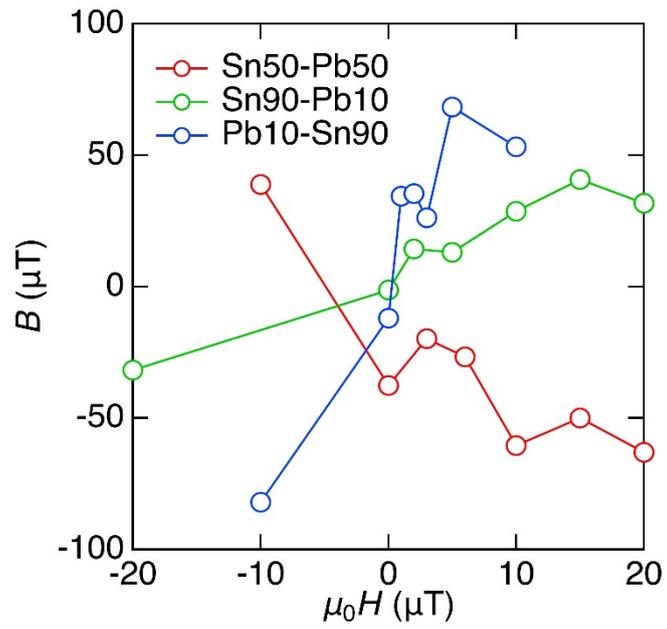

Extended Data Fig. 4 FC field dependence of the total flux sum for the Sn50-Pb50 (red), Sn90-Pb10 (green), and Sn10-Pb90 (blue) solders. The total flux sums were derived from the SSM images shown in Figures S1, S3, and S4 in the Supplementary Information.

Supplementary Information

Figures S1 and S2 present all the scanning SQUID microscopy (SSM) images of Sn50-Pb50 at the zero magnetic field after field cooling (FC) to 5.0 K, where SQUID represents the superconducting quantum interference device. The scanning range is 300 μm × 300 μm in Figure S1 and 200 μm × 200 μm in Figure S2. Immediately after zero-field cooling (ZFC), as shown in Figure S1 (0 μT in Figure S1), the magnetic flux is trapped on the solder surface because of the geometric magnetic field. Increasing the FC field enhances the contrast between the regions of positive and negative trapped fluxes and localised flux-trapping regions begin to emerge. Figure S1 includes an image of the solder after −10 μT FC. Comparing this with the SSM image obtained after +10 μT FC, the red and blue regions in the images are inverted, which demonstrates that magnetic flux trapping in Sn50-Pb50 solder is influenced by the applied external field.

Figure S3 shows all SSM images of the Sn90-Pb10 solder at the zero magnetic field after FC to 4.5 K with a scan range of 300 μm × 300 μm. The contrast between the positive and negative flux regions became more distinct with an increase in the FC field. However, unlike Sn50-Pb50, no flux localisation was observed. Further, Figure S3 includes an SSM image obtained after −10 μT FC, which showed a polarity inversion of the trapped flux when compared to the +10 μT FC image, thereby indicating that trapped flux polarity in Sn90-Pb10 depended on the external field direction.

Figure S4 shows all SSM images of the Sn10-Pb90 solder at a zero magnetic field after FC to 5.0 K. The scan range in Figure S5 is 300 μm × 300 μm. The contrast between the regions of positive and negative trapped magnetic fluxes becomes distinct with an increase in the FC field. Sn10-Pb90 exhibits characteristics that differ from those of the other solders. For example, in the SSM image after 10 μT FC, a localised magnetic flux region appears in the lower right, whereas no such region is observed in the upper left. Figure S5 includes a magnetic image obtained after −10 μT FC. Compared to the image obtained after +10 μT FC, the red and blue regions are inverted.

Figure S5 shows all magneto-optical (MO) images obtained in this experiment. The MO images were acquired immediately after ZFC to 2 K, followed by a magnetic field sequence of 0 mT, → 80 mT → -80 mT → 70 mT in 10 mT increments. The magnetic flux $B$ ($B = 4\pi M + H$) at each magnetic field is shown in Figure 4(a) in the main text. All images are normalised to highlight the magnetisation distribution. The image $I(H, 2.0\ K)$ is subtracted from $I(H, 8.0\ K)$ at a certain $H$, and the result is divided by $I(H, 2.0\ K)$; i.e. $I(H, 2.0\ K)−I(H, 8.0\ K))/I(H\ 2.0\ K)$. The MO image is generally uniform immediately after ZFC. A blue region indicating perfect diamagnetism emerges when a 10 mT magnetic field is applied. Magnetic flux begins to penetrate from the solder edges with an increase in the field, thereby forming a branched pattern. At 80 mT, the entire solder transitions to the normal state, and the MO image becomes uniform. During the magnetic field reduction, trapped magnetic flux regions with branch pattern appear near the solder edges. At −80 mT, the image again becomes uniform. In the subsequent increase in field, negative magnetic flux regions and branched patterns reappear.

Figure S6 illustrates the temperature ($T$) dependence of magnetisation ($M$) in the ZFC and MO images. The MO images are normalised similarly to those in Figure S5, using $I(H, 2.0\,\text{K})-I(H, 8.0\,\text{K}))/I(H, 2.0\,\text{K})$. As shown in Figure S8(b), an external magnetic field of 10 mT was applied to enhance the diamagnetic signal. The $M$–$T$ curve in Figure S6(a) shows that $M$ increases and approaches $M = 0$ near $T_c^{Pb}$ with an increase in temperature, where $T_c^{Pb}$ represents the superconducting transition temperature of Pb. The MO image at 2.0 K shows a negative flux region because of perfect diamagnetism. The negative magnetic flux region diminishes with an increase in temperature, and branched patterns became evident near the solder edges.

Figure S7 presents the $M$–$T$ curve and MO images at the zero magnetic field after 150 mT FC. The $M$–$T$ curve in Figure S7(a) shows positive magnetisation, which indicates flux trapping. $M$ decreases and approaches $M = 0$ near $T_c^{Pb}$ with an increase in temperature. In Figure S7(b), the trapped flux and branched patterns are clearly visible. With increasing temperature, the magnetic flux near the edges gradually escapes from the solder edges.

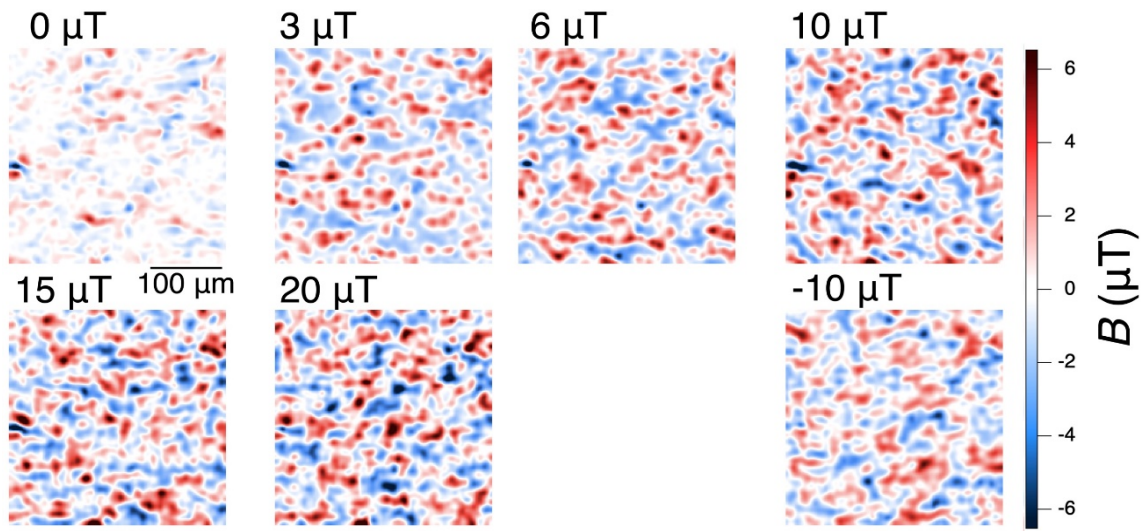

Fig. S1 SSM images at the zero magnetic field after FC to 5.0 K for the Sn50-Pb50 solder. The scanning range is 300 μm × 300 μm.

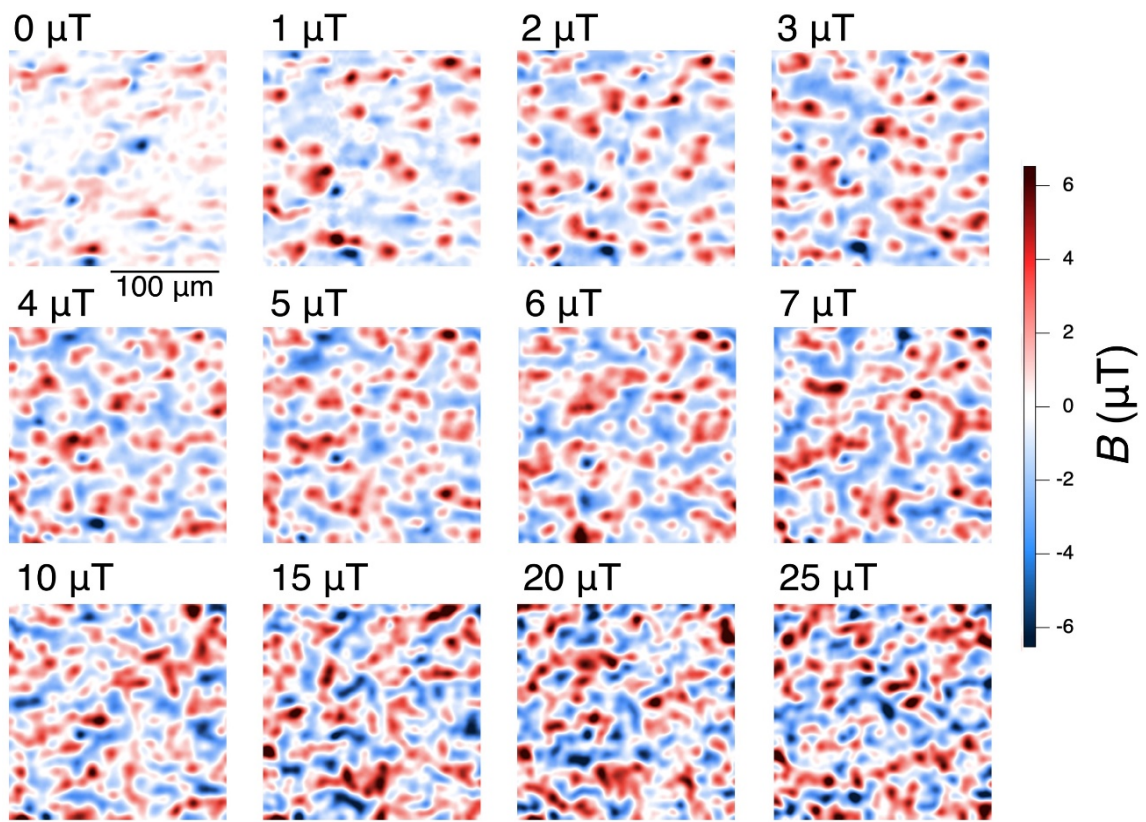

Fig. S2 SSM images at zero magnetic field after FC to 5.0 K in various initial magnetic fields for the Sn50-Pb50 solder. The scanning range is 200 μm × 200 μm.

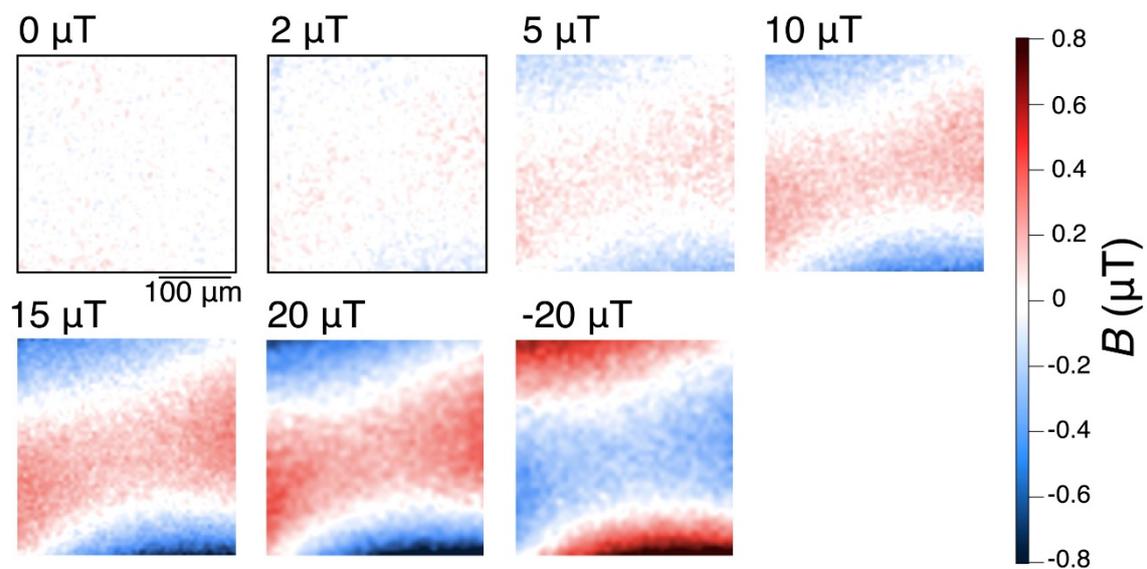

Fig. S3 SSM images at zero magnetic field after FC to 4.5 K for the Sn90-Pb10 solder. The scanning range is 300 μm × 300 μm.

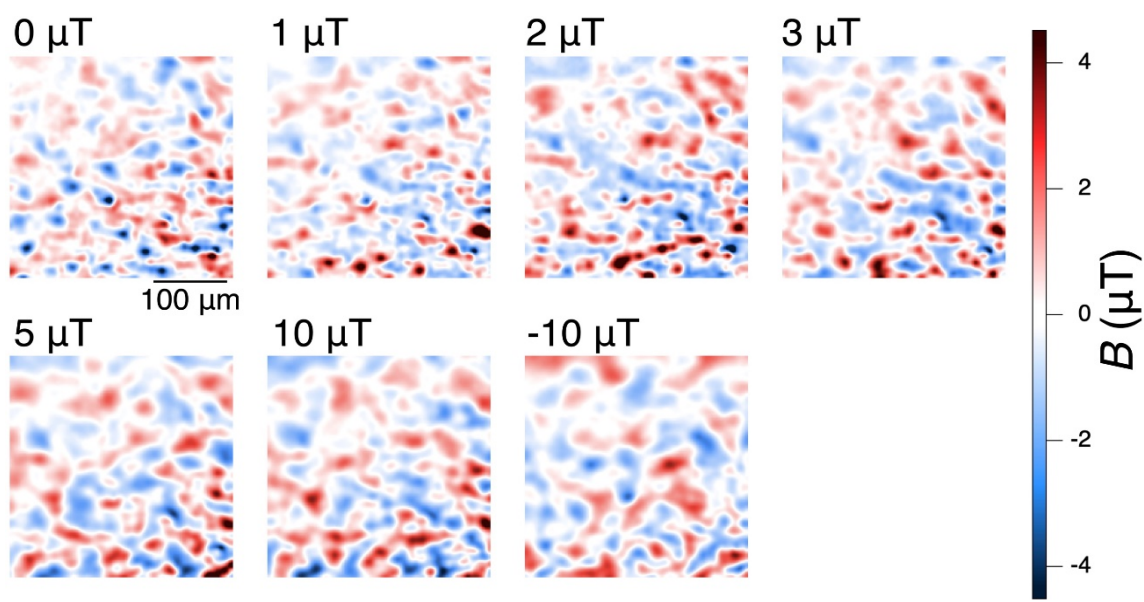

Fig. S4 SSM images at zero magnetic field after FC to 5.0 K for the Sn10-Pb90 solder. The scanning range is 300 μm × 300 μm.

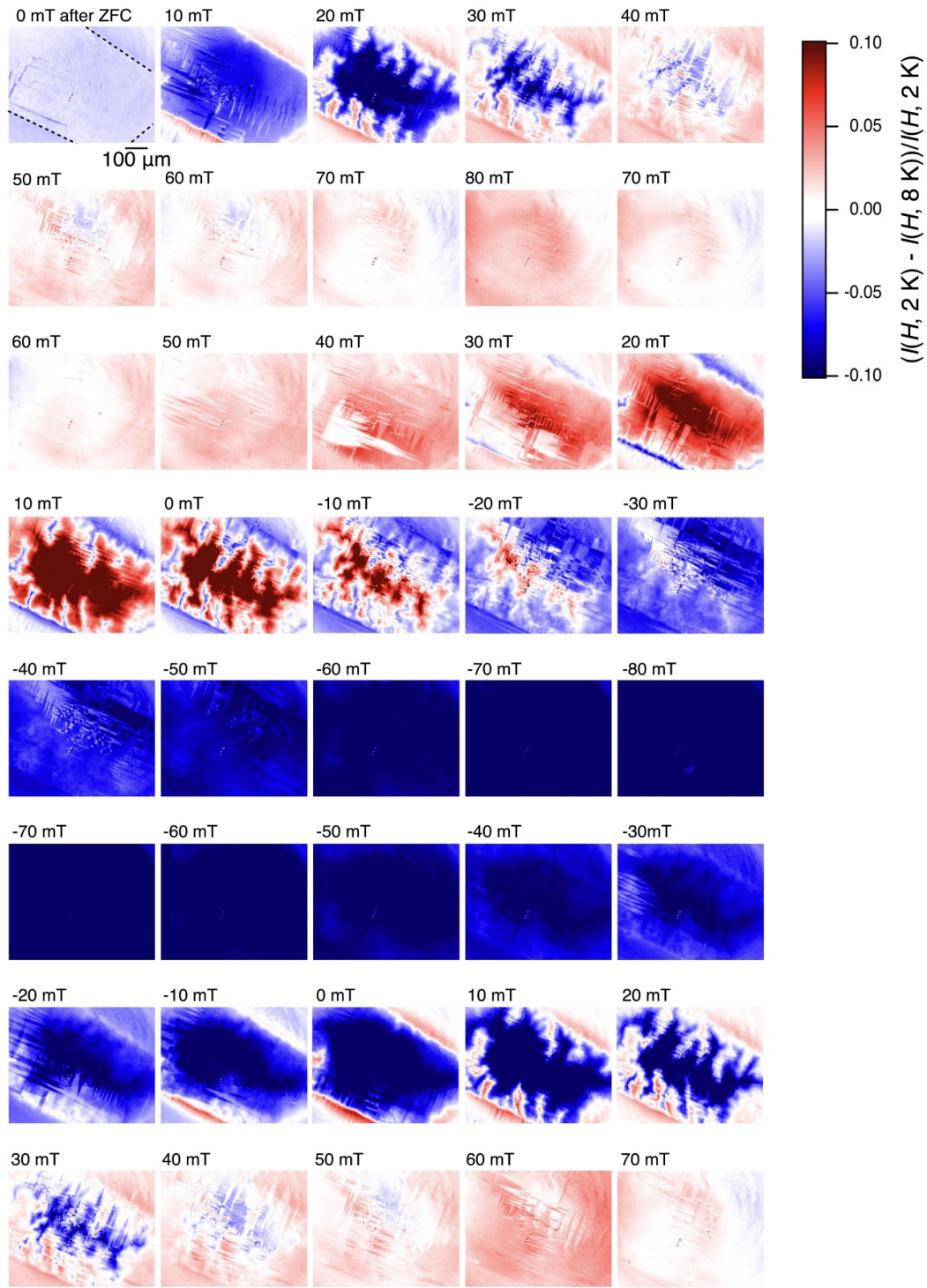

Fig. S5 MO images show the evolution of the magnetic field after ZFC to 2.0 K. The upper row corresponds to the increase in the magnetic field from 0 mT to 80 mT, whereas the middle and lower rows show the reverse process, where the magnetic field decreases from 80 mT to −80 mT and then returns to 80 mT. The colour scale represents the relative change in magnetic flux distribution, normalised to the field at 2.0 K.

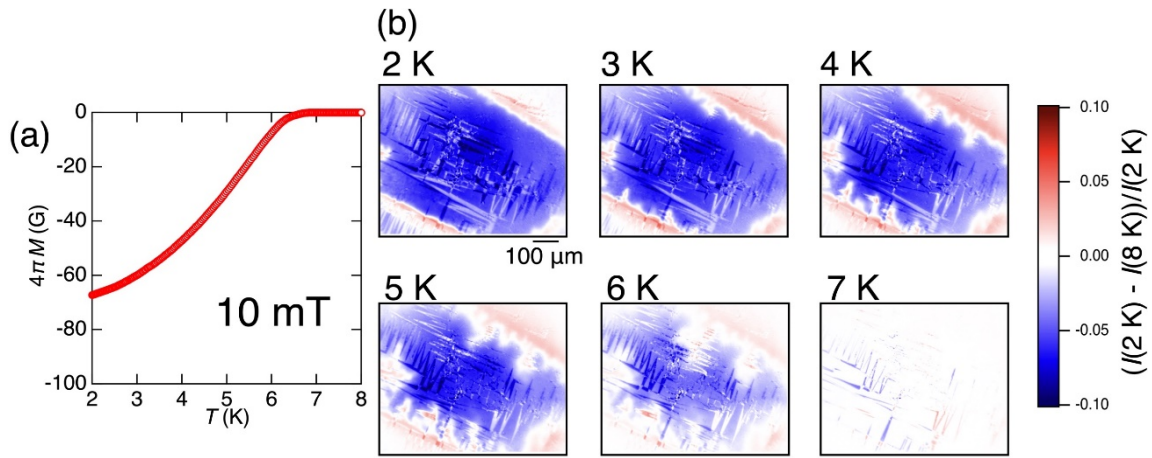

Fig. S6 Temperature-dependent magnetisation and flux distribution in the Sn50-Pb50 solder. (a) *M–T* curve under a 10 mT applied field in the ZFC condition. (b) MO images corresponding to the magnetisation and temperature values in (a), which illustrates the evolution of the magnetic flux distribution with an increase in temperature from 2.0 K to 7.0 K.

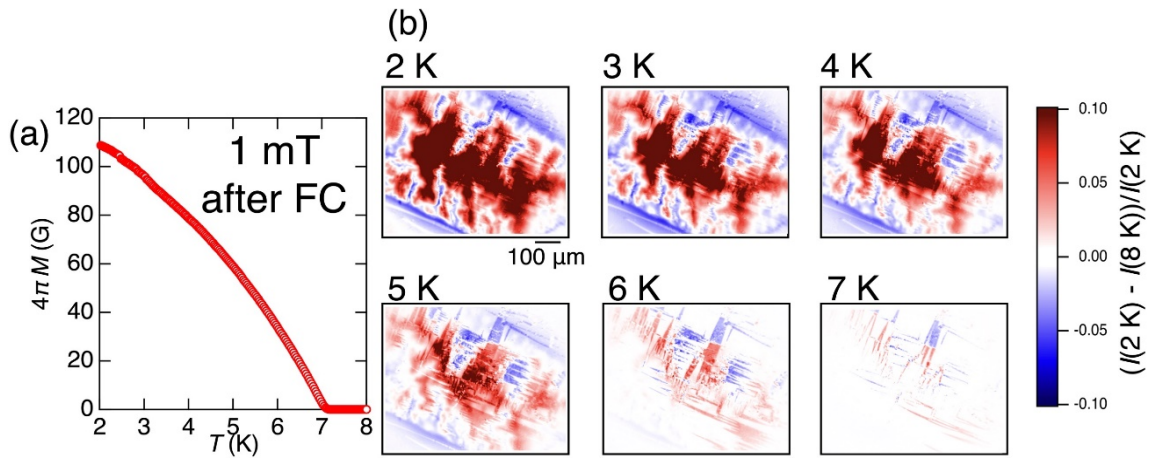

Fig. S7 Temperature-dependent magnetisation and flux distribution in the Sn50-Pb50 solder. (a) The *M–T* curve measured under a 10 mT applied field after 150 mT FC. (b) MO images corresponding to the magnetisation and temperature values in (a), which shows the evolution of the magnetic flux distribution with an increase in the temperature from 2.0 K to 7.0 K.